\documentclass[conference,10pt]{IEEEtran}
\usepackage{graphicx,amsmath,times,setspace,bm}
\IEEEoverridecommandlockouts
\usepackage{cite}
\usepackage{array}
\usepackage{subfig}
\usepackage[ruled,vlined]{algorithm2e}
\usepackage[T1]{fontenc}
\usepackage[mathscr]{eucal}
\usepackage{mathptmx}
\usepackage{multirow}
\DontPrintSemicolon

\usepackage{balance}
\usepackage{amssymb}
\usepackage{xcolor}

\let\mathcal\relax
\DeclareMathAlphabet\mathcal{OMS} {cmsy}{b}{n}
\SetMathAlphabet \mathcal{normal}{OMS}{cmsy}{m}{n}
\DeclareMathAlphabet\mathbcal{OMS} {cmsy}{b}{n}

\DeclareMathOperator*{\argmax}{\arg\!\max}

\begin{document}

\title{\huge Application-Layer Coding with Intermittent Feedback Under Delay and Duty-Cycle~Constraints
\thanks{This work has been supported by the K-project DeSSnet (Dependable, secure and time-aware sensor networks), which is funded within the context of COMET -- Competence Centers for Excellent Technologies by the Austrian Ministry for Transport, Innovation and Technology (BMVIT), the Federal Ministry for Digital and Economic Affairs (BMDW), and the federal states of Styria and Carinthia; the COMET program is conducted by the Austrian Research Promotion Agency~(FFG).}}
\author{
 \IEEEauthorblockN{Siddhartha S. Borkotoky\IEEEauthorrefmark{1,}\IEEEauthorrefmark{2}, Udo Schilcher\IEEEauthorrefmark{1,}\IEEEauthorrefmark{3}, and Christian Raffelsberger\IEEEauthorrefmark{1}}
   \IEEEauthorblockA{\IEEEauthorrefmark{1}Lakeside Labs GmbH,
     9020 Klagenfurt, Austria}
    \IEEEauthorblockA{\IEEEauthorrefmark{2}Indian Institute of Technology Bhubaneswar, Khordha 752050, India}
    \IEEEauthorblockA{\IEEEauthorrefmark{3}University of Klagenfurt,
    9020 Klagenfurt, Austria}
    (borkotoky@iitbbs.ac.in)
    \vspace*{-2mm}
    }

\maketitle

\begin{abstract}
We propose two application-layer coding schemes for delay-constrained point-to-point packet communications with restrictions on the transmitter's maximum duty-cycle. The schemes operate over GF(2) and utilize intermittently available receiver feedback for erasure correction. Applications that will benefit from the proposed schemes include wireless sensor networks in which energy-constrained sensors must deliver readings to a gateway within a deadline. Simulation results for independent Bernoulli erasure channels, Gilbert-Elliott channels, and Long Range (LoRa) communications demonstrate orders-of-magnitude reductions in the delivery failure rate as compared to feedback-assisted repetition redundancy and a blind coding scheme that does not utilize feedback.
\end{abstract}
\begin{IEEEkeywords}
Application-layer coding, delay-sensitivity, duty cycling, LoRa, wireless sensor networks.
\end{IEEEkeywords}

\renewcommand{\baselinestretch}{0.9}
\small\normalsize

\section{Introduction}
\label{intro}

Certain wireless networks are subject to limits on the transmitters' maximum duty cycle. Such limits could be due to regulatory restrictions in certain frequency bands or could be enforced to restrict the transmission times of energy-constrained devices. Duty-cycle constraints limit the use of receiver feedback in multipoint-to-point communications as the receiver cannot acknowledge all receptions without violating the constraint when the number of transmitters is large. They also restrict the maximum number of retransmissions a sender can make and the amount of redundancy it can transmit for forward error correction. 

The focus of this paper is on duty-cycle-constrained point-to-point transfer of delay-sensitive information. In the scenarios of interest, an information symbol (a collection of information bits) remains of interest to the receiver only for a limited amount of time after it is generated. Our work is motivated by use cases from a wireless sensor network in which sensors periodically transmit readings to a gateway using the LoRa technology~\cite{RKS17} over the 868 MHz unlicensed frequency band, for which the maximum allowed duty cycle is 1\,\% in the European Union~\cite{AVT17}. Due to this restriction, the gateway acknowledges only some of the transmissions. The sensors, likewise, limit their packet sizes to conserve energy and meet duty-cycle regulations. Each measurement must reach the gateway within a specified deadline. To improve communications reliability under such circumstances, we propose two application-layer coding schemes for erasure correction.

Although motivated by LoRa use cases, our proposed schemes can be implemented with any physical-layer technology. In our schemes, a transmitted packet contains the current information symbol, and possibly some past information symbols  as well as some coded symbols (random linear combinations of information symbols).  The schemes employ cumulative feedback that conveys the identity of the oldest unexpired-but-undelivered information symbol. Instead of blindly inserting coded symbols into every packet, current and past feedback is used to choose between transmitting coded symbols and retransmitting past information symbols.  When coding occurs, the feedback is used to determine the number of information symbols to combine to produce a coded symbol as well as the set of information symbols from which the combined symbols are randomly chosen.  The computational complexity is kept low by coding only at certain instances and restricting all coding operations to GF(2). In our design, we account for the possibility that the sender may not obtain receiver feedback after every transmission. 



\section{Related Work}
\label{related_work}
Well-known application-layer codes such as LT codes~\cite{Lub02} and Raptor codes~\cite{Sho06}  operate over large blocks of  information symbols. They have strong erasure-correction capabilities, but large decoding delays render them unsuitable for delay-constrained applications.

Coding over small blocks of information is investigated in~\cite{DKF13} under the assumption of perfect feedback. In~\cite{FEL19}, a coding scheme over a large finite field (GF(256)) for delay-sensitive multimedia applications is proposed. In contrast, we seek computationally simpler strategies that operate in GF(2). Unlike in~\cite{DKF13} and~\cite{FEL19}, where each coded symbol is sent in a separate packet, we consider the transmission of coded symbols and information symbols in the same packet. By avoiding the extra packet overhead in this manner, lower energy consumption and smaller duty cycles are achieved. For example, transmitting a \mbox{3-byte} sensor measurement  followed by a 3-byte coded measurement in two separate packets using LoRa results in 67\,\% higher duty-cycle and energy expenditure, compared to sending the measurement and the coded measurement in the same packet~\cite{Sem13}. Additionally, in some situations, coded symbols can be transmitted without increasing the packet duration. For example, the  duration of a LoRa packet remains constant for payload sizes of 1 byte through 4 bytes~\cite{Sem13}. Therefore, a 1-byte measurement can be accompanied by up to three coded measurements of 1 byte each at no extra cost in terms of duty-cycling and energy expenditure.

The approach closest to ours is that of~\cite{MRP17}, where each packet contains the current information symbol and some coded symbols that are bitwise XORs of randomly selected past information symbols. However, the scheme assumes complete absence of feedback, and hence cannot benefit from any feedback availability. Furthermore, in contrast to the insertion of coded symbols into each packet in~\cite{MRP17}, our methods perform coding only when deemed necessary.

\section{System Model}
\label{system_model}

\begin{figure}
\centering
\includegraphics[scale=0.38,bb=-50 400 520 470]{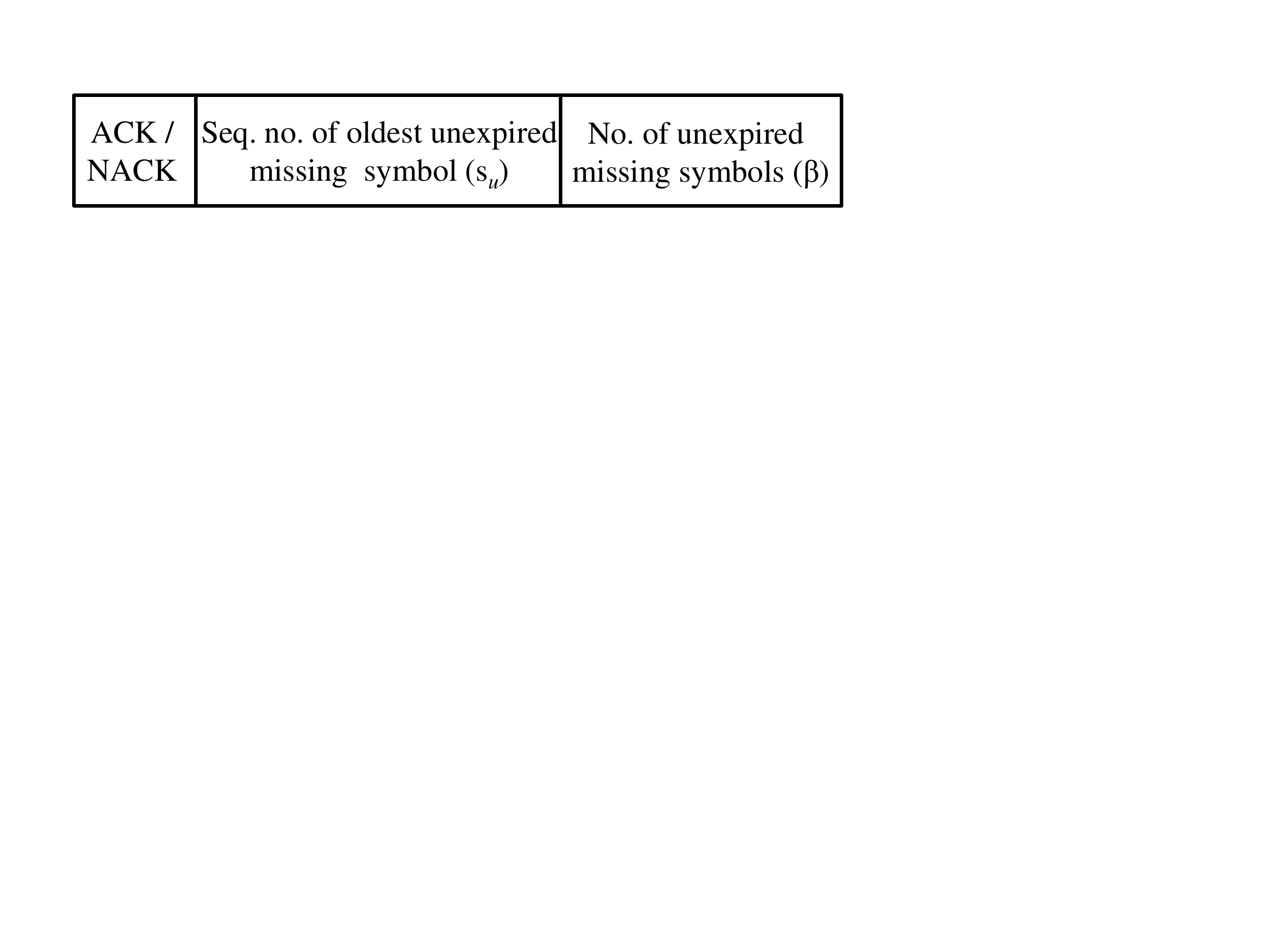}
\setlength{\belowcaptionskip}{-5pt}
\caption{Feedback structure.}
\label{feedback_structure}
\end{figure}
\renewcommand{\baselinestretch}{0.9}
\small\normalsize

\begin{figure}
\centering
\includegraphics[scale=0.38,bb=90 300 520 520]{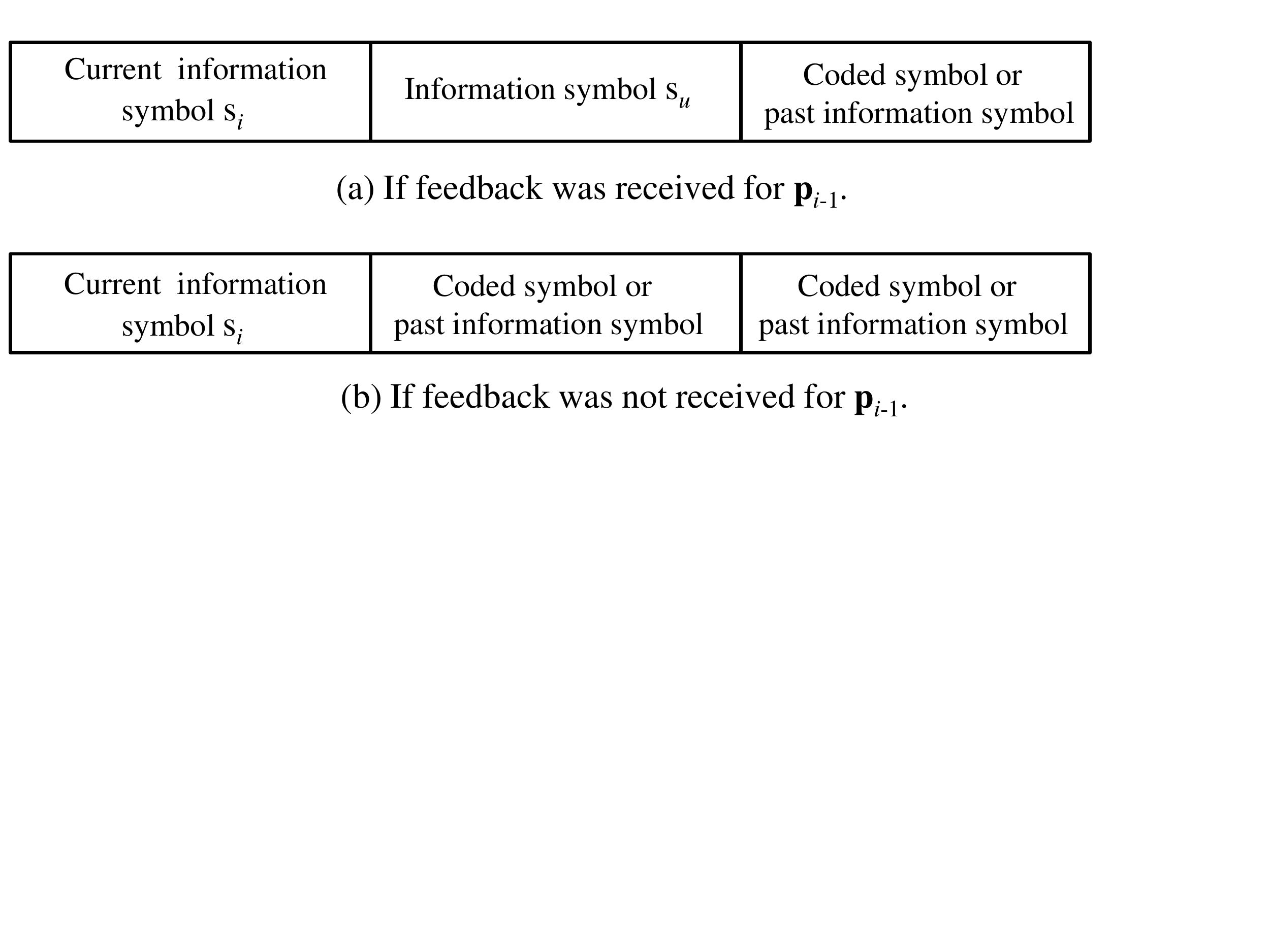}
\setlength{\belowcaptionskip}{-10pt}
\caption{Payload structure for packet $\mathbf{p}_i$ ($b = 3$).}
\label{payload_structure}
\end{figure}
\renewcommand{\baselinestretch}{0.9}
\small\normalsize

\begin{table} [t]
\caption{Summary of key notation}
\begin{tabular}{ c | c  }
    \hline
    Notation & Description \\ \hline
    $\mathrm{s}_i$ &  Information symbol no. $i$\\ \hline
    $\mathbf{p}_i$ &  Packet no. $i$\\ \hline
    $b$ & Maximum number of symbols \\
    & a packet can carry \\ \hline
    $\mathbf{q}_i[k]$ &  $k$-th symbol in the payload of $\mathbf{p}_i$ \\
    &($1\leq k \leq b$) \\  \hline
    $u$ &  Sequence number of oldest unexpired \\
     & and undelivered information symbol ($\mathrm{s}_u$)\\  \hline
    $\beta$ & Number of unexpired information symbols\\
    & not yet delivered \\ \hline
    $\mathbf{c}_d(\mathbf{v})$ & A coded symbol produced by XORing \\
    & $d$ randomly chosen elements from vector $\mathbf{v}$, \\
    & where $d$ is determined using~\eqref{degree_selection}\\  \hline
    $\mathbf{c}_D(\mathbf{v})$ & A coded symbol produced by XORing \\
    & $D$ randomly chosen entries of vector $\mathbf{v}$, \\
    & where $D$ is a uniform random variable\\ \hline
    $\mathbf{v}[k]$ & $k$-th entry of vector $\mathbf{v}$ \\ \hline
    $\mathbf{v} \! \smallsetminus \! \{x\}$ & Set containing all entries of $\mathbf{v}$ except $x$ \\ \hline
\end{tabular}
\label{Notation}
\end{table}

An \emph{information symbol} is a collection of information bits (e.g., the binary representation of a sensor measurement). Every information symbol triggers a packet transmission. Let $\delta_{\max}$ denote the application's \emph{delay tolerance}, the time after which an information symbol expires (is no longer of interest to the receiver).

A \emph{coded symbol }is a linear combination (bitwise XOR) of information symbols selected at random from a set of information symbols referred to as the  \emph{coding set}. The \emph{degree} of the coded symbol is the number of information symbols combined and can vary from symbol to symbol. A coded symbol of degree $d$ results in the recovery of an information symbol at the receiver if \mbox{$d -1$} out of the $d$ XORed information symbols have already been delivered.  An information symbol is \emph{delivered} if the receiver either receives a packet containing the uncoded symbol or is able to recover it from a coded symbol.

An information symbol and a coded symbol are both \mbox{$l$-bits} long. Let $l_p$ be the maximum number of bits in a packet's payload, as determined by duty-cycle and energy constraints. Thus, $b = l_p/l$ is the maximum number of symbols that can be included in the packet.

Following a transmission, the sender receives a feedback packet with a feedback reception probability (FRP) $p_{\mathrm{feedback}}$. As shown in Fig.~\ref{feedback_structure}, the feedback packet contains a one-bit ACK or NACK that indicates whether the transmission was received correctly, the sequence number $u$ of the oldest undelivered and unexpired information symbol, and the total number $\beta$ of undelivered and unexpired information symbols. If all unexpired symbols up to and including $\mathrm{s}_j$ in the just-transmitted packet $\mathbf{p}_j$ have been delivered, then $u = j + 1$ in the feedback for~$\mathbf{p}_j$. 

The payload structure of the $i$-th packet $\mathbf{p}_i$ is illustrated in Fig.~\ref{payload_structure} for $b = 3$. Packet $\mathbf{p}_i$ includes the information symbol $\mathrm{s}_i$. If an acknowledgement was received for $\mathbf{p}_{i-1}$ and $u \neq i$, then $\mathbf{p}_i$ includes $\mathrm{s}_u$ as well. The space for the remaining \mbox{$b-2$} symbols carry either past information symbols or coded symbols, as determined by the coding algorithms.

A summary of the key notation used in the paper is given in Table~\ref{Notation}.

\section{Coding Schemes}
\label{coding_schemes}
The two proposed coding schemes differ in their respective coding sets and in the determination of the degrees of the coded packets. In \emph{windowed coding}, the coding set is a continuous window of information symbols starting from the oldest unexpired and undelivered symbol until the most recently transmitted one. In \emph{selective coding}, some symbols from the aforementioned window are excluded based on receiver feedback.
Descriptions and pseudocodes for the algorithms are given below.

\subsection{Windowed Coding}
The payload of packet $\mathbf{p}_i$ depends on whether a feedback was received for the previous packet  $\mathbf{p}_{i-1}$:
\subsubsection{Feedback received for $\mathbf{p}_{i-1}$}
From the feedback, the sender knows that the set of sent but undelivered (and unexpired) information symbols is a subset of the set of entries in the vector \mbox{$\mathbf{w}_i = \{\mathrm{s}_u, \mathrm{s}_{u+1}, \ldots, \mathrm{s}_{i-1} \}$}. Note that $\mathbf{w}_i$ has $i-u$ entries, and if $\beta = i -u$, no symbol in $\mathbf{w}_i$ has been delivered. There are the following possibilities:

\vspace{1mm} {(a) $u = i$}:
All information symbols up to this point have been delivered. Hence, $\mathbf{p}_i$ contains only $\mathrm{s}_i$ in its payload.

\vspace{1mm} {(b) $u < i$, $i - u \leq b - 1$, $\beta \geq 1$}:
All entries of $\mathbf{w}_i$, which form a superset of all undelivered information symbols, can be included in the packet. Hence, $\mathbf{p}_i$ contains $\mathrm{s}_i$ followed by the elements of  $\mathbf{w}$.

\vspace{1mm} {(c) $u < i$, $i - u > b - 1$, $1 < \beta = i - u$}:
None of the information symbols in $\mathbf{w}_i$ have been delivered, but there is not enough space in $\mathbf{p}_i$ to include all of them. Hence, $\mathbf{p}_i$ contains $\mathrm{s}_i$ and the first $b-1$ elements of $\mathbf{w}_i$ (i.e., \mbox{$\{\mathrm{s}_j: u \leq j \leq u + b - 2\}$}).

\vspace{1mm} {(d) $u < i$, $i - u > b - 1$, $1 < \beta < i - u$}:
Here, $\beta$ out of the $i-u$ entries of $\mathbf{w}_i$ are undelivered; but the sender does not know which ones (except for $\mathrm{s}_u$), nor can it include the entire vector $\mathbf{w}_i$ in $\mathbf{p}_i$. We leverage coding in this situation. Packet $\mathbf{p}_i$ contains $\mathrm{s}_i$, $\mathrm{s}_u$, and $b-2$ coded symbols produced by random linear combinations of elements from $\mathbf{a}_i = \mathbf{w}_i  \smallsetminus \{ \mathrm{s}_u\}$. The degree of each coded symbol is $d(i-u-1,\beta-1)$, where
\begin{equation} \label{degree_selection}
d(x,y) = \argmax_{d'}  \frac{y {x - y \choose d' - 1}}{{x \choose d'}}\mbox{ for }x > y > 0\:.
\end{equation}
The degree is chosen to maximize the probability that exactly one of the XORed information symbols belongs to the set of $\beta - 1$ undelivered symbols, thus maximizing the probability that the coded symbol results in the recovery of one information symbol. Note that there are $\beta$ undelivered information symbols, but the symbol $\mathrm{s}_u$ included in $\mathbf{p}_i$ reduces that number to $\beta - 1$.

The maximization according to~\eqref{degree_selection} is only an approximation if there are multiple coded symbols in the packet. A joint maximization of the individual degrees is possible, but is not employed to keep the complexity low.

\subsubsection{Feedback not received for $\mathbf{p}_{i-1}$}
Define {$\mathbf{b}_i = \{\mathrm{s}_{i-1},\mathrm{s}_{i-2},\ldots,\mathrm{s}_{i-z}\}$}, where \mbox{$z = \min\{ i - u_l, i - u_{\max}\}$}, $u_l$ is the sequence number included in the most recent feedback, and $u_{\max}$ is the oldest unexpired information symbol at time $i$. There are two possibilities:

\vspace{1mm} {(a) $z \leq b-1$}:
The entire vector $\mathbf{b}_i$, whose entries form a superset of all undelivered information symbols, can be included in the packet. Hence, $\mathbf{p}_i$ contains $\mathrm{s}_i$ followed by the entries of $\mathbf{b}_i$.

\vspace{1mm} {(b) $z > b-1$}:
This is similar to situation (d) above, but since the current number of undelivered symbols is unknown, degree selection via~\eqref{degree_selection} is infeasible. Packet $\mathbf{p}_i$ contains $\mathrm{s}_i$ followed by $b-1$ coded symbols. A coded symbol is the XOR of $D$ information symbols chosen at random from $\mathbf{b}_i$, where $D$ is a uniform random variable in the range $[1, z]$.

\begin{algorithm}[t]
  \nl $\mathbf{q}_i[1] = \mathrm{s}_i$\;
  \nl \If{\text{Feedback received for} $\mathbf{p}_{i-1}$}{
        \If{$u < i$}{
            $\mathbf{q}_i[2] = \mathbf{w}_i[1]$  \;
            \If{$\beta > 1$} {
                \If {$i - u \leq b - 1$}   {
                    $\mathbf{q}_i[k] = \mathbf{w}_i[k-1]$,  $3 \leq k \leq i - u + 1$\;
                }
                \ElseIf {$i - u > b - 1$ and $\beta = i - u$} {
                    $\mathbf{q}_i[k] = \mathbf{w}_i[k-1]$,  $3 \leq k \leq b$\;
                }
                \ElseIf {$i - u > b - 1$, $1 < \beta < i - u$} {
                    $\mathbf{q}_i[k] = \mathbf{c}_d(\mathbf{a}_i)$,  $3 \leq k \leq b$\;
                }
            }
        }
  }
  \Else{
    \If {$z \leq b-1$}   {
        $\mathbf{q}_i[k] = \mathbf{b}_i[k]$,  $2 \leq k \leq z + 1$\;
    }
    \Else {
        $\mathbf{q}_i[k] = \mathbf{c}_D(\mathbf{b}_i)$, $2 \leq k \leq b$\;
    }

  }
\caption{Windowed coding}
\label{coding_algo1}
\end{algorithm}

\begin{algorithm}[t]
  \nl $\mathbf{q}_i[1] = \mathrm{s}_i$\;
  \nl \If{\text{Feedback received for} $\mathbf{p}_{i-1}$}{
        \If{$\beta \geq 1$}{
            $\mathbf{q}_i[2] = \mathrm{s}_u$  \;
            \If {$\beta > 1$} {
                \If {$n = \beta$}   {
                    $\mathbf{q}_i[k] = \mathbf{n}_i[k - 2]$,  $3 \leq k \leq \max\{n+1, b\}$\;
                }
                \ElseIf {$b - 1 \geq n > \beta$} {
                    $\mathbf{q}_i[k] = \mathbf{n}_i[k - 2]$,  $3 \leq k \leq n + 1$\;
                }
                \ElseIf {$n > \beta$ and $n > b - 1$} {
                    $\mathbf{q}_i[k] = \mathbf{c}_d(\mathbf{n}_i)$, $3 \leq k \leq b$ \;
                }
            }
        }
  }
  \Else{
    \If {$n < b - 1$}   {
        $\mathbf{q}_i[k] = \mathbf{m}_i[k]$, $2 \leq k \leq n+1$\;
    }
    \Else {
        $\mathbf{q}_i[k] = \mathbf{c}_D(\mathbf{m}_i)$, $2 \leq k \leq b$\;
    }

  }
\caption{Selective coding}
\label{coding_algo2}
\end{algorithm}

\subsection{Selective Coding} \vspace{-0.03in}

The sender maintains a list of the unexpired information symbols for which an ACK has not been received and whose sequence numbers are equal to or greater than $u_l$, the sequence number included in the most recent feedback. Let $\mathbf{m}_i$ be the vector of such symbols at time $i$, and let $n$ denote the number of entries in $\mathbf{m}_i$. The payload of $\mathbf{p}_i$ is constructed as follows:

\subsubsection{Feedback received for $\mathbf{p}_{i-1}$}

There are the following possibilities:

\vspace{1mm} {(a) $\beta = 0$}:
The receiver has received all information symbols up to this point. Hence, $\mathbf{p}_i$ contains only $\mathrm{s}_i$.

\vspace{1mm} {(b) $\beta = 1$}:
There is only one undelivered symbol at the receiver, which is $\mathrm{s}_u$. Hence, $\mathbf{p}_i$ contains $\mathrm{s}_i$ and $\mathrm{s}_u$.

\vspace{1mm} {(c) $\beta > 1$, $n = \beta$}:
The sender knows which symbols are undelivered. Here, $\mathbf{p}_i$ contains $\mathrm{s}_i$, $\mathrm{s}_u$, and $\max\{n-1, b-2\}$  information symbols from the vector $\mathbf{n}_i = \mathbf{m}_i \! \smallsetminus \! \{\mathrm{s}_u\}$.

\vspace{1mm} {(d) $\beta > 1$, $b - 1 \geq n > \beta$}:
The sender does not know which symbols are undelivered, but all potentially undelivered symbols can be included in $\mathbf{p}_i$. Here, $\mathbf{p}_i$ contains $\mathrm{s}_i$ followed by the elements of $\mathbf{m}_i$.

\vspace{1mm} {(e) $\beta > 1$, $n > \beta$, $n > b - 1$}:
The sender knows a set of $n$ symbols, out of which $\beta$ are undelivered (although the sender does not know which ones), and not all $n$ symbols fit in the packet. In this case, $\mathbf{p}_i$ contains $\mathrm{s}_i$, $\mathrm{s}_u$, and $b - 2$ coded symbols produced by XORing $d$ randomly chosen information symbols out of the $n - 1$ symbols in \mbox{$\mathbf{n}_i = \mathbf{m}_i \! \smallsetminus \! \{\mathrm{s}_u\}$}. The degree $d$ is determined according to~\eqref{degree_selection} with $x=n-1$ and $y=\beta-1$.

\subsubsection{Feedback not received for $\mathbf{p}_{i-1}$}
There are two possibilities:

\vspace{1mm} {(a) $n \leq b - 1$}:
All potentially undelivered information symbols can be included in a packet. Hence, $\mathbf{p}_i$ contains $\mathrm{s}_i$ followed by the information symbols in $\mathrm{m}_i$.

\vspace{1mm} {(b) $n > b - 1$}:
Similar to situation (e), but the current number of undelivered symbols is unknown. Packet $\mathbf{p}_i$ contains $\mathrm{s}_i$ followed by $b - 1$ coded symbols, each produced from the coding set comprising the elements of $\mathbf{m}_i$  and with a degree chosen uniformly at random from $[1,n]$.

For both coding schemes, the maximization in~\eqref{degree_selection} can be performed using a look-up table containing the values of $d(x,y)$ for all valid $(x,y)$ pairs. Since the applications are delay constrained, the number of pairs is not large. If there can be a maximum of $q$ unexpired packets at any given instant, the look-up table has $q(q+1)/2$ entries. For our numerical results, $q \leq 16$, resulting in at most 136 table entries.

\section{Performance Evaluation}
\label{results}
\subsection{Simulation Setup}
We are interested in the fraction of information symbols that are not delivered before their expiration. This performance measure is called the {\it delivery failure rate} (DFR). We simulate periodic packet transmissions. Unless stated otherwise, an information symbol expires $\delta_{\max} = 16$ packet intervals after it is first transmitted. For each data point in the numerical results, enough transmissions are simulated to obtain at least 100 delivery failures. Results are provided for three types of channels:

{\it a) Independent Bernoulli erasure channel}.

{\it b) Gilbert-Elliott channel}: Using a two-state Markov chain,  packets transmitted during the ``good'' state are delivered and packets in the ``bad'' state are lost. The state is fixed during a packet transmission but may change from packet to packet. The transition probability from ``good'' to ``bad'' (``bad'' to ``good'') is denoted by $p_{\mathrm{gb}}$ ($p_{\mathrm{bg}}$).

{\it c) LoRa packet-erasure model}: The tool LoRaSim~\cite{LoRaSim} is used to simulate LoRa communications in an industrial environment. The spreading factor~\cite{Sem13} is 10 and each packet has a duration of 13 LoRa symbols, which corresponds to a payload of 4 bytes or lower~\cite{Sem13}.  Packet losses occur due to independent fading (Nakagami-$m$  with $m = 2.5$) and interference from other LoRa transmitters. The nodes are simulated as points on a two-dimensional plane. The receiver is at the origin; the sender is at $x = y = 36$~m. There are up to 200 other transmitters in the vicinity; the x and y-coordinates of each such transmitter are uniform random variables in the range \mbox{[30 m, 42 m].} The pathloss exponent is 4, which results in short ranges.

For each setup, the arrival of feedback is an independent Bernoulli random process. Following a transmission, the sender receives feedback with probability $p_{\mathrm{feedback}}$. The non-arrival of feedback may be due to disturbances in the feedback channel, or due to the inability of the receiver to send a feedback caused by duty-cycle restrictions. We do not attempt to derive a stochastic characterization of the feedback reception process but employ the simple Bernoulli model instead.

\subsection{Benchmark Schemes}
Two benchmark schemes are simulated:

1) A \emph{repetition-redundancy scheme} that utilizes feedback but does not perform XORing of symbols. Packet $\mathbf{p}_i$ contains the current information symbol $\mathrm{s}_i$ and, if a feedback was received for $\mathbf{p}_{i-1}$, the symbol $\mathrm{s}_u$ whose sequence number was included in the feedback. The remainder of the payload carries the most recent information symbols that have not been acknowledged yet.

2) A \emph{blind coding scheme} (similar to that of~\cite{MRP17}) that does not utilize feedback; each packet contains the current information symbol followed by $b-1$ coded symbols formed by XORing randomly chosen unexpired information symbols. The degree of a coded symbol is $\delta_{\max}/2$. As shown in~\cite{MRP17}, the degree requires careful adjustment for optimal performance, but such adjustments are infeasible unless the channel model is known in advance.

\subsection{Simulation Results}

\begin{figure}
    \subfloat[$b=2$, $p_{\mathrm{feedback}} = 0.25$.]{\label{Bernoulli_b2_p25}\includegraphics[scale=0.29,bb=-100 0 770 440]{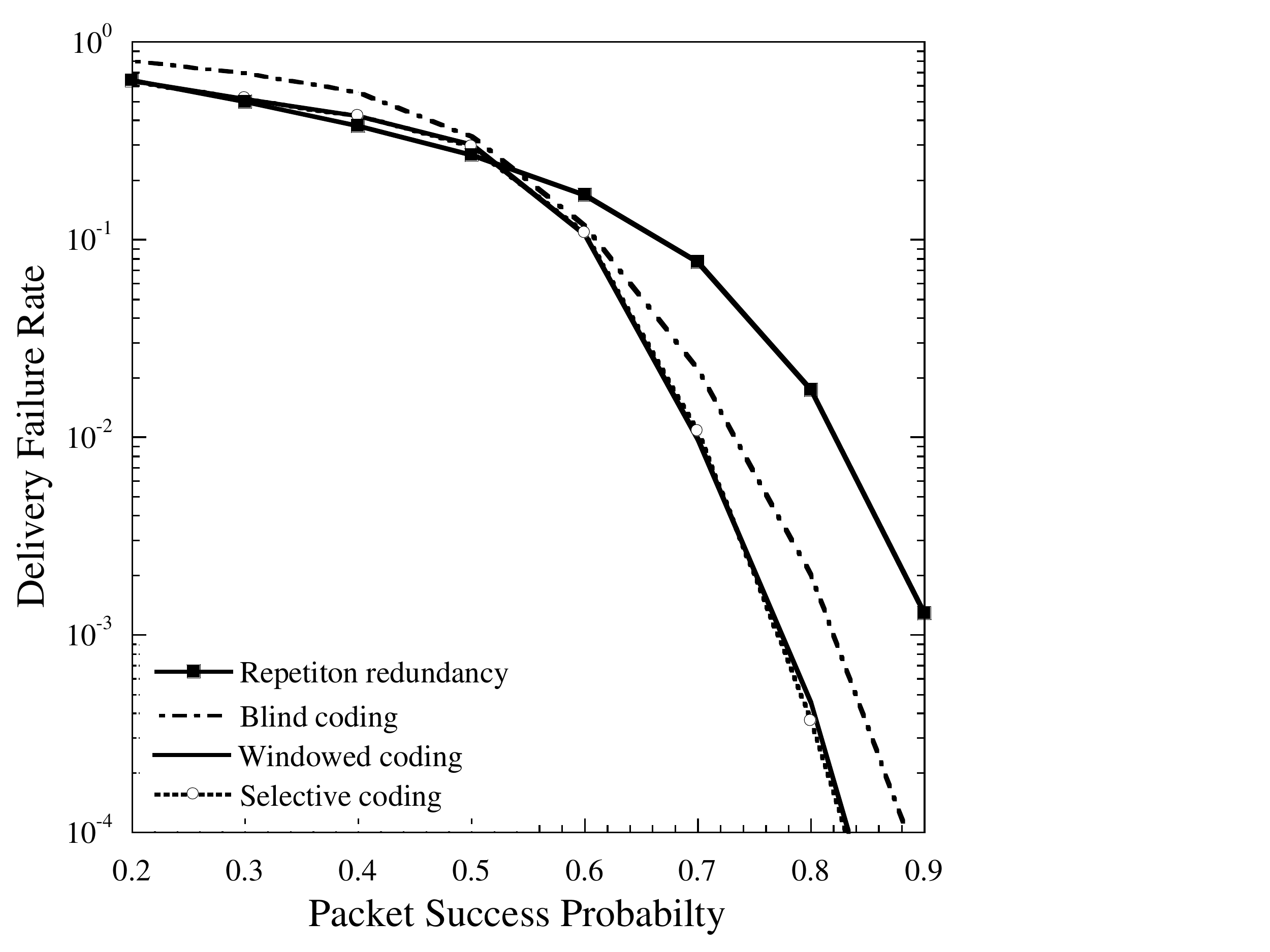}} \\
    \subfloat[$b=3$, $p_{\mathrm{feedback}} = 0.75$.]{\label{Bernoulli_b3_p75}\includegraphics[scale=0.29,bb=-100 0 770 510]{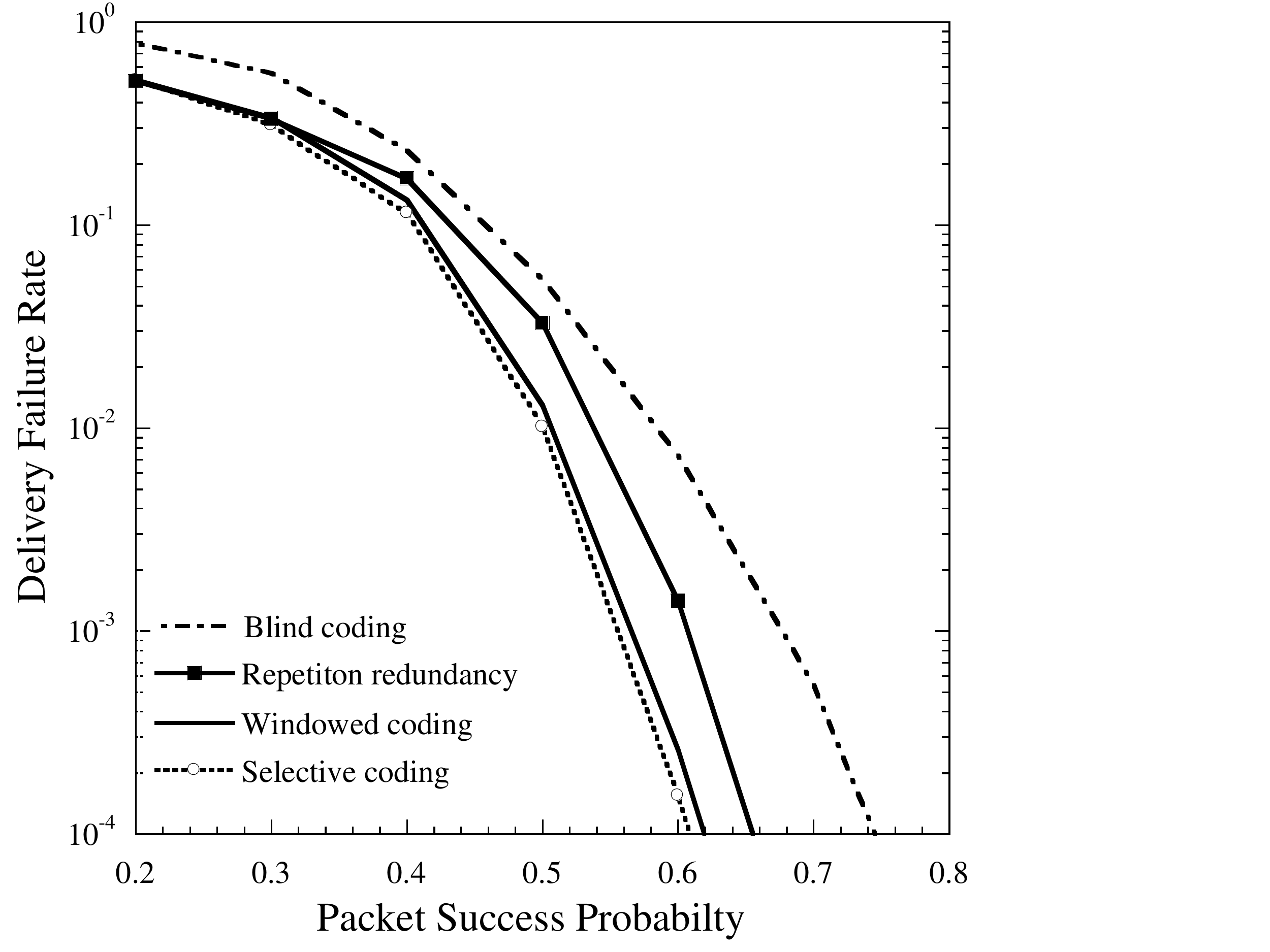}}
    \setlength{\belowcaptionskip}{-2pt}
    \caption{Delivery performance on an independent Bernoulli erasure channel.}
    \label{Bernoulli}
\end{figure}

Fig.~\ref{Bernoulli} shows the DFR over an independent Bernoulli channel for two sets of values of $b$ and $p_{\mathrm{feedback}}$. Fig.~\ref{Bernoulli_b2_p25} shows that the proposed algorithms outperform repetition redundancy for packet success probabilities exceeding 0.5, providing up two orders-of-magnitude improvement. Up to an order-of-magnitude improvement over the blind coding scheme is also seen. The two proposed schemes provide similar performance. In Fig.~\ref{Bernoulli_b3_p75}, the DFR of each of the four schemes is lower than in Fig.~\ref{Bernoulli_b2_p25}. This is due to the additional redundancy per packet and, except for the blind coding scheme, more frequent availability of feedback. In this scenario, selective coding outperforms windowed coding. Also, unlike Fig.~\ref{Bernoulli_b2_p25}, the blind coding scheme performs much worse than the other three schemes due to its inability to exploit the greater availability of feedback.

\begin{figure}
    \includegraphics[scale=0.28,bb=-100 50 770 510]{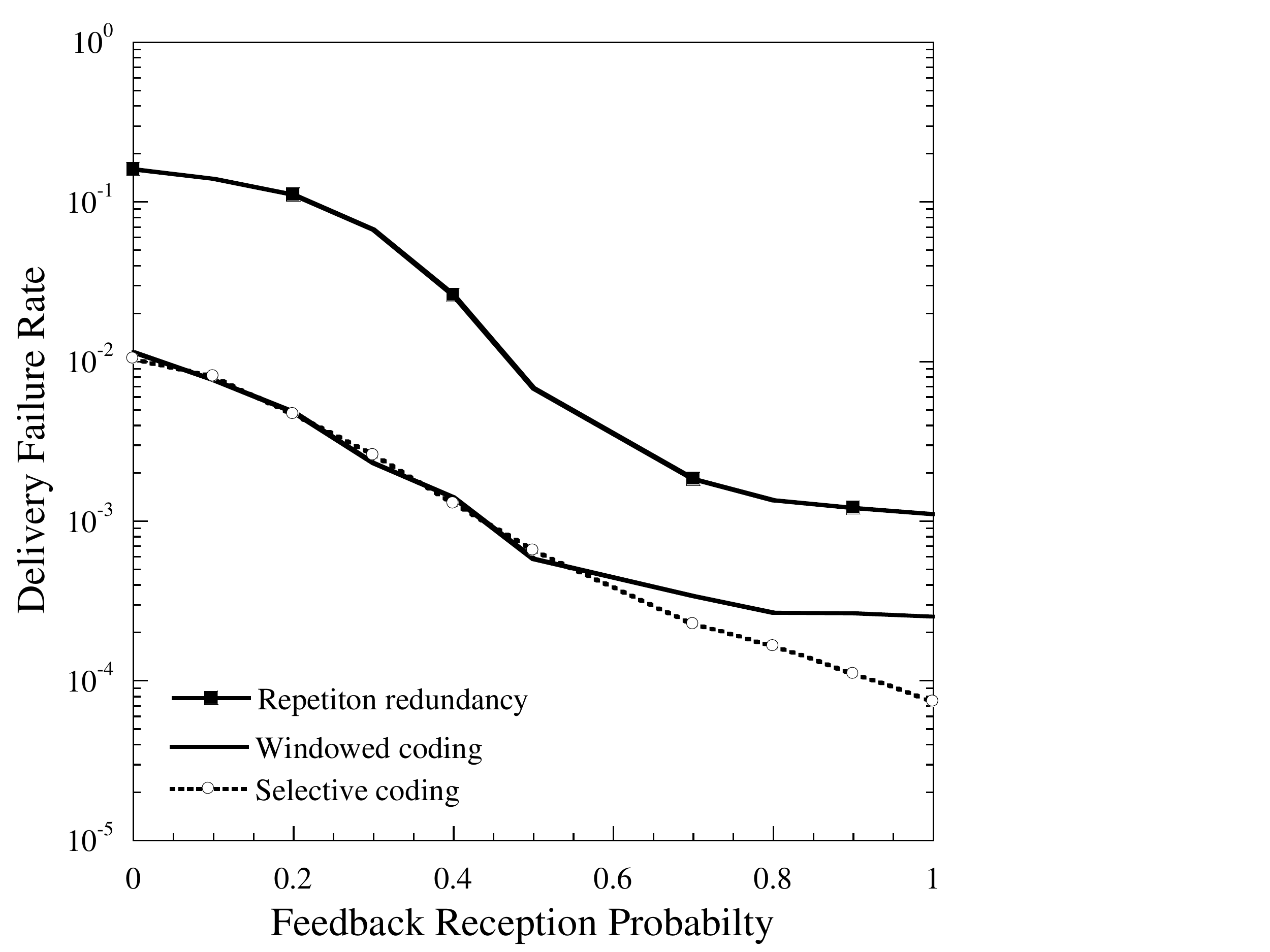}
    \setlength{\belowcaptionskip}{-2pt}
    \caption{Delivery performance as a function of FRP.}
    \label{Bernoulli_FRP}
\end{figure}

\begin{figure}
    \subfloat[$b=2$, $p_{\mathrm{feedback}} = 0.3$.]{\label{GE_b3_p30}\includegraphics[scale=0.29,bb=-100 0 770 440]{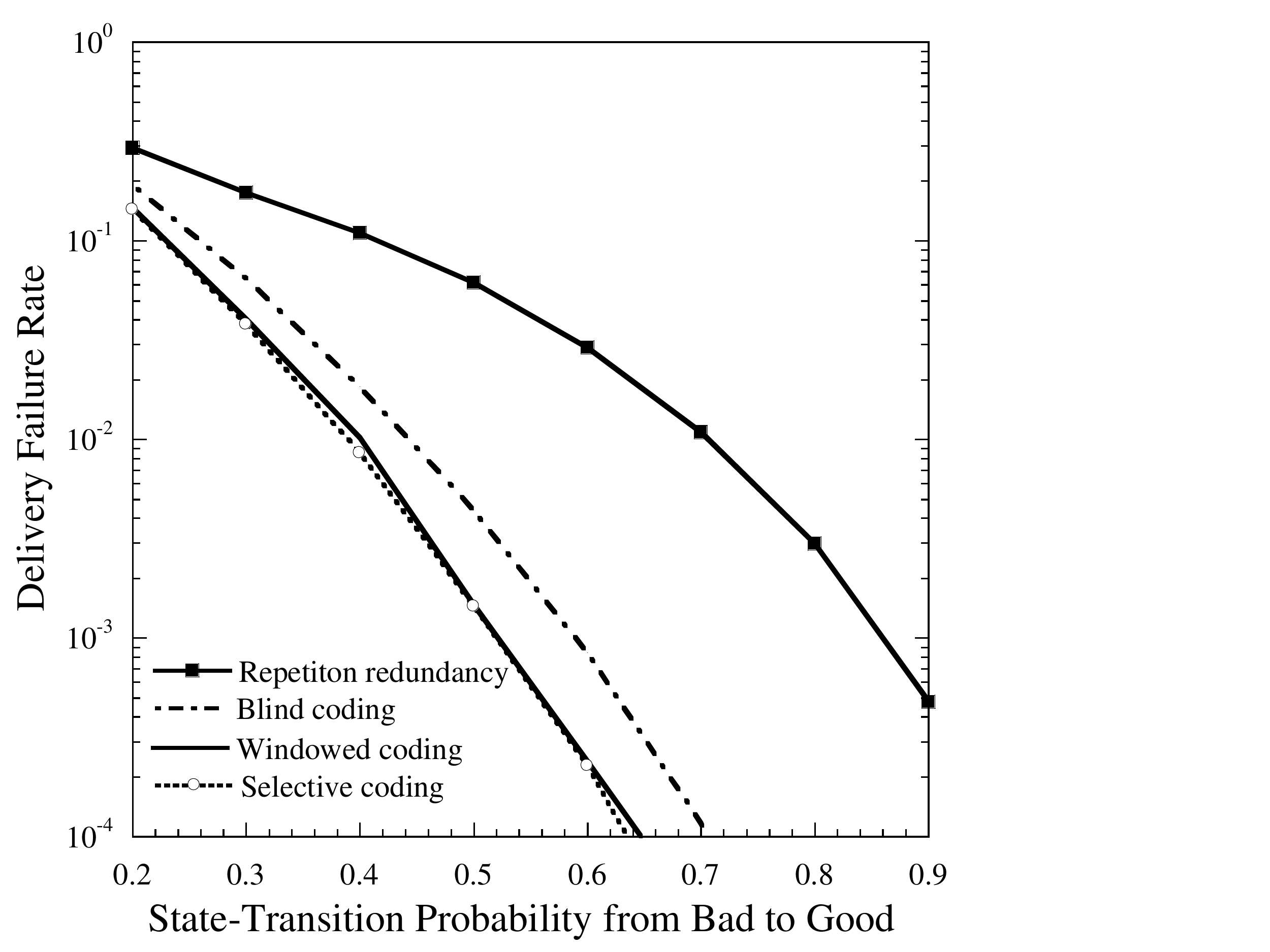}} \\
    \subfloat[$b=3$, $p_{\mathrm{feedback}} = 0.9$.]{\label{GE_b3_p90}\includegraphics[scale=0.29,bb=-100 0 770 510]{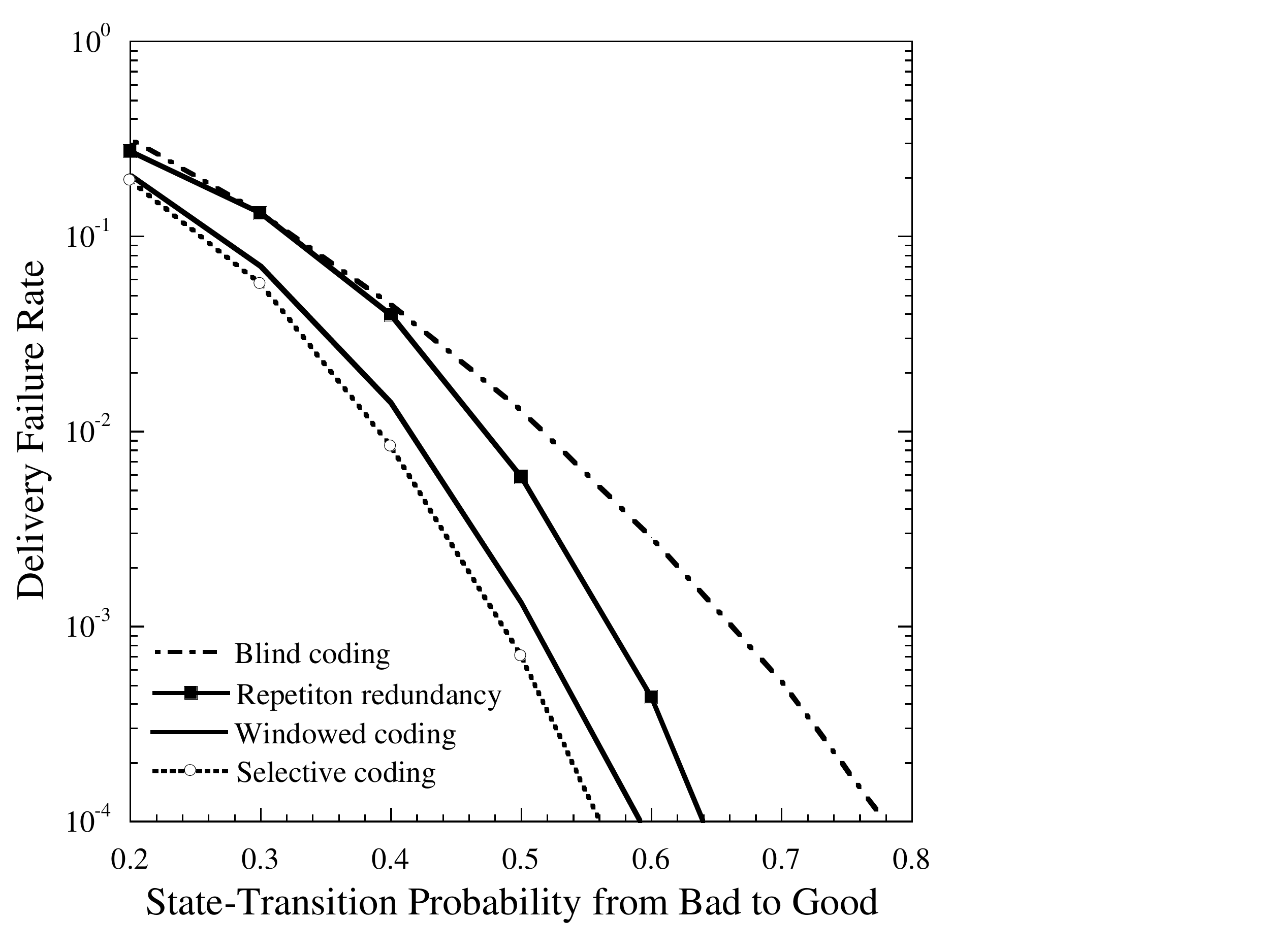}}
    \setlength{\belowcaptionskip}{-2pt}
    \caption{Delivery performance on a Gilbert-Elliott channel.}
    \label{GE}
\end{figure}

Fig.~\ref{Bernoulli_FRP} plots the DFR against FRP. The link from sender to receiver is an independent Bernoulli erasure channel with success probability 0.6. The proposed algorithms outperform repetition redundancy for any value of the FRP. For low FRPs, the two proposed algorithms give approximately the same performance. However, selective coding outperforms windowed coding when the FRP is high and, unlike windowed coding and repetition redundancy, its performance does not saturate at high FRP.

Results for the Gilbert-Elliott channel are shown in Fig.~\ref{GE}. The transition probability $p_{\mathrm{gb}}$ is fixed at 0.2 while $p_{\mathrm{bg}}$ is varied. The graphs exhibit similar patterns to those for the independent Bernoulli erasure channel, but the performance improvement due to the proposed algorithms is higher than for the independent Bernoulli channels.

\begin{figure}
    \includegraphics[scale=0.29,bb=-100 50 770 540]{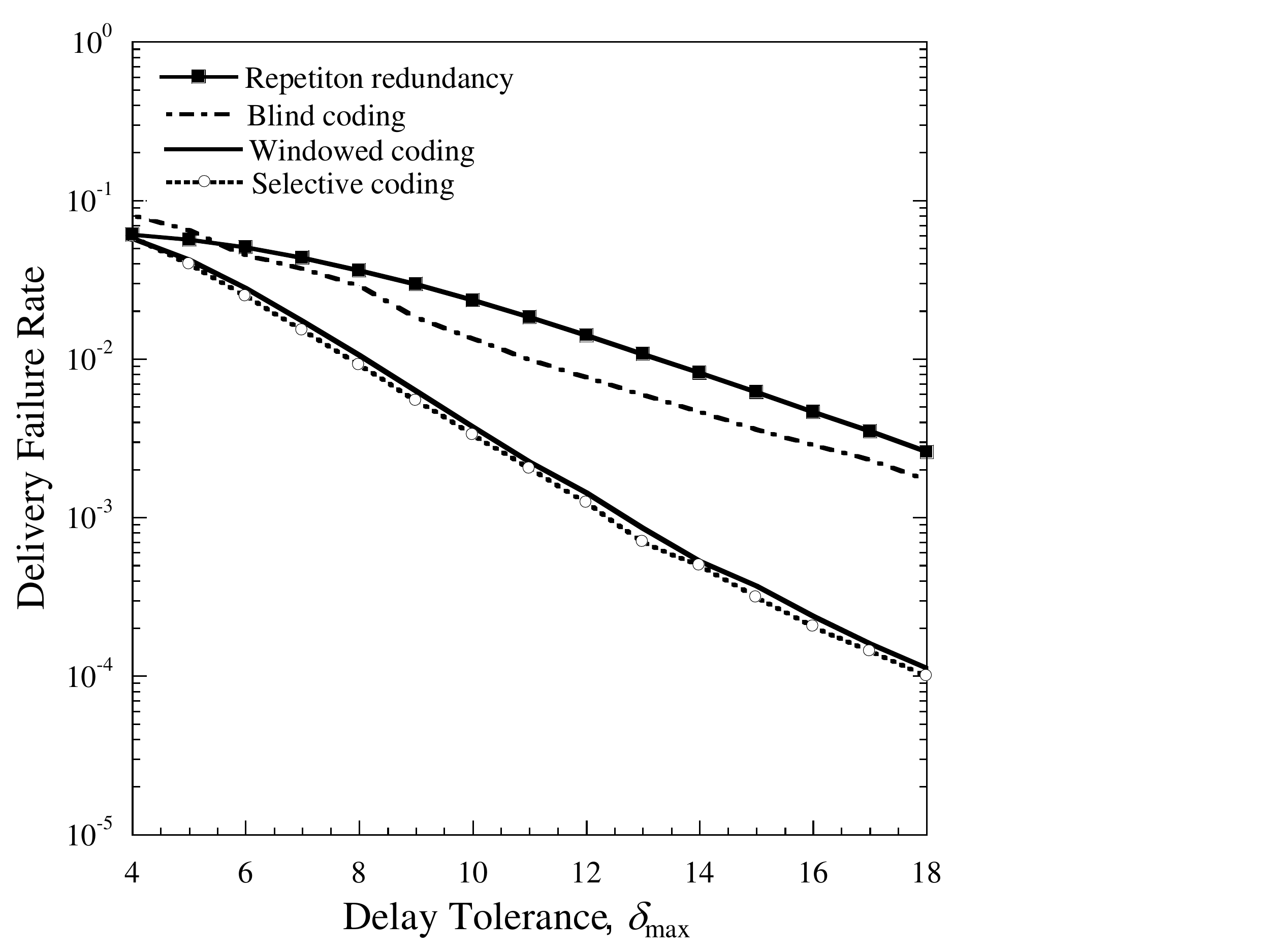}
    \setlength{\belowcaptionskip}{-0pt}
    \caption{Delivery performance versus delay tolerance.}
    \label{GE_dmax}
\end{figure}

Fig.~\ref{GE_dmax} shows the delivery performance against the delay tolerance. Recall that an information symbol expires $\delta_{\max}$ packet intervals after its first transmission. The results are shown for a Gilbert-Elliott channel with \mbox{$p_\mathrm{gb}=0.3$}, \mbox{$p_\mathrm{bg}=0.6$}, and \mbox{$p_{\mathrm{feedback}}=0.5$}.  The number of symbols per packet is $b = 3$. We observe that higher delay tolerance results in better performance for each scheme, which is due to more opportunities to correct an erasure. Higher delay tolerance also leads to a greater performance advantage of the proposed algorithms over the benchmark schemes.

\begin{figure}
    \includegraphics[scale=0.30,bb=-100 20 770 510]{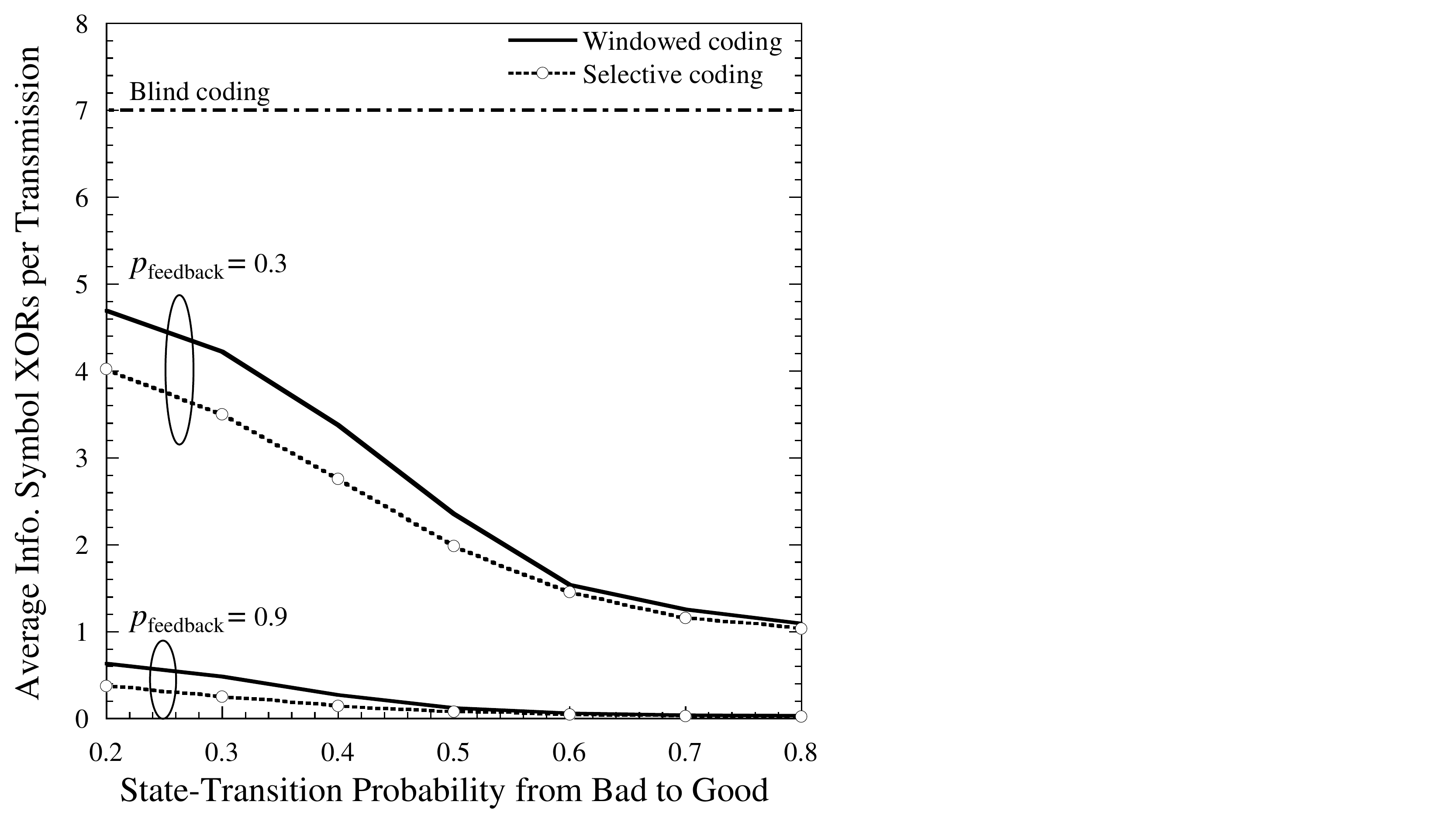}
    \setlength{\belowcaptionskip}{-0pt}
    \caption{Average number of information symbols XORed per transmission.}
    \label{GE_Complexity}
\end{figure}

To assess the computational overhead due to coding, we plot in Fig.~\ref{GE_Complexity} the average number of symbols XORed per transmitted packet for the Gilbert-Elliott channel with $p_\mathrm{gb} =0.2$ and for two different feedback reception probabilities. We observe that higher feedback availability reduces the number of symbol  XORs for the proposed schemes by providing the sender with more information about the set of potentially undelivered packets. Similarly, the number decreases with better channel conditions owing to fewer undelivered information symbols.  For each value of $p_{\mathrm{feedback}}$, selective coding incurs lower complexity than windowed coding by further winnowing the coding set. Due to its use of a fixed degree, the number of XORs in the blind coding scheme is constant.

\begin{figure}
    \subfloat[$b=2$.]{\label{LoRa_b2}\includegraphics[scale=0.31,bb=-100 0 770 510]{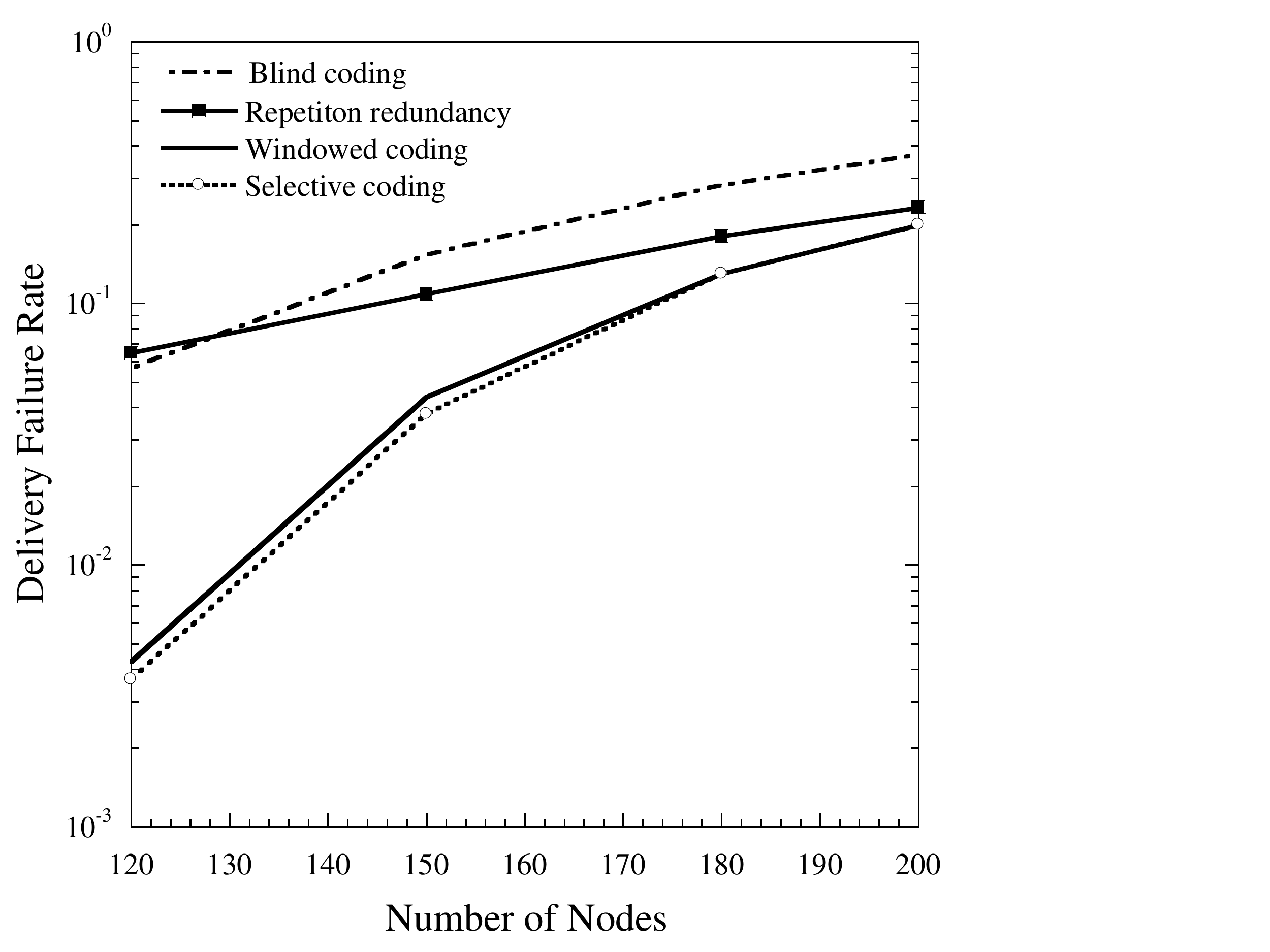}} \\
    \subfloat[$b=4$.]{\label{LoRa_b4}\includegraphics[scale=0.31,bb=-100 0 770 507]{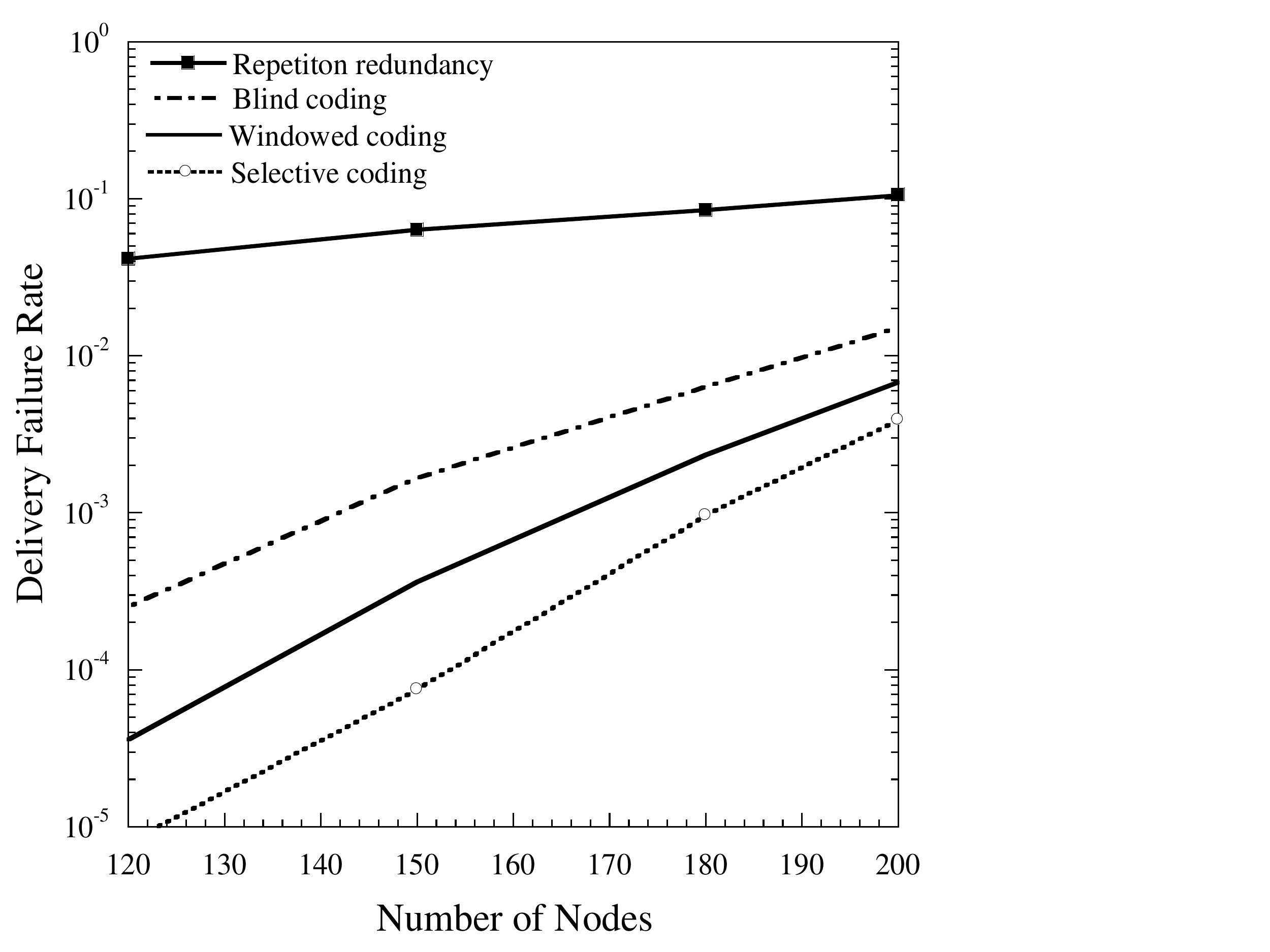}}
    \setlength{\belowcaptionskip}{-0pt}
    \caption{Delivery performance for LoRa.}
    \label{LoRa}
\end{figure}

Fig.~\ref{LoRa} shows the delivery performance for LoRa communications between a sensor and a gateway. Each sensor measurement is 1 byte long. For the LoRa parameters employed, the packet duration is constant for payload sizes of 1--4 bytes~\cite{Sem13}. Therefore, up to 3 symbols in addition to the current information symbol (i.e., up to $b = 4$) can be included without increasing the duty cycle and transmission energy. The FRP is 0.5. The delivery performance is plotted against the number of nearby nodes. The DFR increases with the number of nodes due to higher interference. As before, the proposed coding schemes outperform the benchmark schemes. For $b=4$, we observe that the performance advantage of selective coding over windowed coding is higher than in the other scenarios considered so far.

\vspace*{-1mm}
\section{Conclusion}
\label{conclusion} \vspace*{0mm}
We proposed two application-layer coding schemes -- windowed coding and selective coding -- for delay-constrained communications with limited receiver feedback. The schemes are effective tools to improve communications reliability in various delay-sensitive and duty-cycle-constrained settings, such as those encountered in wireless sensor networks. As with any coding technique, there is some increase in computations relative to uncoded redundancy transmissions; however, the added complexity is much lower than for blind coding, and decreases with increasing packet success probability and feedback reception probability. Selective coding provides superior  delivery performance and lower computational complexity than windowed coding at the cost of some increase in storage requirements (due to the need to maintain a list of unacknowledged symbols). The designer can choose one of the schemes depending on target delivery performance and the transmitting node's computational capability and storage capacity.

Future work includes the following: (a) A computationally simple method for joint optimization of the degrees of the coded symbols when there are multiple coded symbols per packet. (b) Adaptation of the proposed schemes to the point-to-multipoint scenario (e.g., a sensor sending data to multiple gateways).
\bibliographystyle{IEEEtran}
\bibliography{references}

\end{document}